\documentclass[12pt,a4paper]{article}

\usepackage{amsmath}
\usepackage{amssymb}
\usepackage{mathrsfs}
\usepackage{graphicx}
\usepackage{dsfont}
\usepackage[pdftex,colorlinks=true,linkcolor=black,citecolor=black,urlcolor=black]{hyperref}
\usepackage{enumerate}

\numberwithin{equation}{section}

\def\eqn#1{ \begin{eqnarray} #1 \end{eqnarray} }
\def\eq#1 { \begin{equation} #1 \end{equation} }
\def\eqn#1{ \begin{eqnarray} #1 \end{eqnarray} }

\def\a{\alpha}
\def\b{\beta}
\def\g{\gamma}
\def\e{\epsilon}
\def\w{\omega}

\def\psib{\overline\psi}

\def\cC{\mathcal{C}}
\def\cL{\mathcal{L}}

\def\cO{\mathcal{O}}

\def\sl2r{SL(2,\mathbb{R})}

\def\ce{\varepsilon}

\newcommand{\lsim}{\mathrel{\hbox{\rlap{\lower.55ex \hbox{$\sim$}} \kern-.3em \raise.4ex \hbox{$<$}}}}
\newcommand{\gsim}{\mathrel{\hbox{\rlap{\lower.55ex \hbox{$\sim$}} \kern-.3em \raise.4ex \hbox{$>$}}}}

\newcommand{\diag}[1]{{\mathrm{diag}\{ #1 \}}}
 \newcommand{\be}{\begin{equation}}
\newcommand{\ee}{\end{equation}}

\newcommand{\vev}[1]{\left< #1 \right>} 

\usepackage{bm}

\begin{document}

\title{\begin{flushright}\vspace{-1in}
       \mbox{\normalsize  EFI-14-34}
       \end{flushright}
       \vskip 20pt
Hall viscosity, spin density, and torsion}

\date{\today}

\author{
Michael Geracie
  \thanks{\href{mailto:golkar@uchicago.edu}
    {mgeracie@uchicago.edu}},~
Siavash Golkar
  \thanks{\href{mailto:golkar@uchicago.edu}
    {golkar@uchicago.edu}},~
   Matthew M. Roberts
   \thanks{\href{mailto:matthewroberts@uchicago.edu}
     {matthewroberts@uchicago.edu}},~
      \\ \\
   {\it \it Kadanoff Center for Theoretical Physics and Enrico Fermi Institute,}\\
   {\it   University of Chicago, 5640 South Ellis Ave., Chicago, IL 60637 USA}
} 

\maketitle

\begin{abstract}
We investigate the relationship between Hall viscosity, spin density and response to geometric torsion. For the most general effective action for relativistic gapped systems, the presence of non-universal terms implies that there is no relationship between torsion response and Hall viscosity. We also consider free relativistic and non-relativistic microscopic actions and again verify the existence of analogous non-universal couplings. Explicit examples demonstrate that torsion response is unrelated to both Hall viscosity and spin density. We also argue that relativistic gapped theories must have vanishing Hall viscosity  in Lorentz invariant vacuums.
\end{abstract}

\section{Introduction}
\label{sec:intro}


Recently, there has been much interest in parity odd transport (also called Hall transport)  properties of condensed matter systems. One particular property that has received attention is the relationship between the Hall viscosity of a system and the ``mean orbital spin per particle''

\eq{
\eta_H = \frac{\vev\ell}2 . \label{eq:hall_spin}
} Initially derived in \cite{Read:2008rn}, it was argued that various trial quantum Hall wavefunctions have this property. This was later proven more rigorously in the general case of gapped or topological phases in \cite{ReadRezayi:2010}. It was argued that the quantity on the right hand side is not always the total angular momentum of the system. More recently, it was shown in \cite{Hoyos:2014lla} that in the absence of a magnetic field, this relation should hold even in the presence of gapless modes if the only gapless modes of the theory are Goldstones. In this case the right hand side was taken to be:
\begin{equation}
	\vev\ell=\frac1V \int \ell(x),\label{eq:read_formula}
\end{equation}
where $\ell(x)$ is the total orbital angular momentum about the point $x$. 

Effective field theories (EFTs) have been shown to describe bulk Hall transport in both non-relativistic \cite{Wen:1992ej,Wen:1992uk,Son:2013} and relativistic \cite{Golkar:2014,Golkar:2014paa} settings. It would be interesting to test the viscosity-spin density relationship in its various forms in the framework of these EFTs. There are two major difficulties in this calculation. First, it is not clear what the quantity on the right hand side of the equation corresponds to in the EFT and second, some of the prescriptions given require the analysis of these EFTs at finite volume which has not yet been looked at in detail.

Here, we will focus on one proposed prescription that does not have these difficulties. It has been conjectured that one can define the spin current as response to torsion \cite{Hughes:2012vg}. Therefore, to test the relationship \eqref{eq:hall_spin}, one needs to generalize the EFT to a geometry with torsion and calculate the torsion response. This procedure was carried out in the case of the EFT for a non-relativistic topological phase \cite{Bradlyn:2014wla}, where it was shown that in generality the relationship defined in such way does not hold. It was then argued that if one were to specialize to the case where the microscopic theory is that of a minimally coupled spinless particle which does not feel torsion, then one should ignore all torsion couplings in the EFT. If one then includes a Wen-Zee term with the torsionful connection, $A \wedge d \omega$, then the equality is restored. The authors of \cite{Bradlyn:2014wla} used this reasoning to argue that \eqref{eq:hall_spin} in fact holds. To us however, this discrimination of including a torsionful Wen-Zee term while at the same time discarding torsion couplings in the rest of the effective action is arbitrary. Starting with a microscopic theory that does not feel torsion implies that the effective theory will not couple to torsion either. Hence the Wen-Zee term can only include the \emph{torsionless} Levi-Civita connection, $A \wedge d \tilde \omega [e]$ which gives a non-zero Hall viscosity while the  torsion response is zero by construction.

In a similar fashion, we will demonstrate that in other examples the torsion response does not give the correct answer in that \eqref{eq:hall_spin} is not satisfied when defining $\vev\ell$ via response to torsion. As we will see, the problem is that there are non-universal couplings to torsion in all cases we consider which spoil the relationship and are not forbidden by symmetry. We will argue this for both the relativistic and non-relativistic case, looking at the free microscopic theory as well as the effective continuum theory in the relativistic case.

A corollary of our study is that the only contribution to Hall viscosity in EFTs describing relativistic gapped systems is through coupling to the Euler current \cite{Golkar:2014,Golkar:2014paa}, which requires a background field to spontaneously break Lorentz invariance. Therefore in a Lorentz invariant vacuum the Hall viscosity must vanish. This is of course consistent with \eqref{eq:hall_spin}.

The outline of this paper is as follows. In section \ref{sec:tors_bg} we briefly review  Riemann-Cartan geometry and its relation to lattice defects. In section \ref{sec:dirac_R} we write down the most general action for a free Dirac fermion and look at its response to torsion. We then move on to analyze the effective field theory of a relativistic quantum Hall system in section \ref{sec:QH_R}. Finally, in section \ref{sec:NR_free} we perform an analysis for non-relativistic fermions along the lines of section \ref{sec:dirac_R}.

\section{Geometry with curvature and torsion}
\label{sec:tors_bg}

\begin{figure}
\centering
\includegraphics[scale=.2]{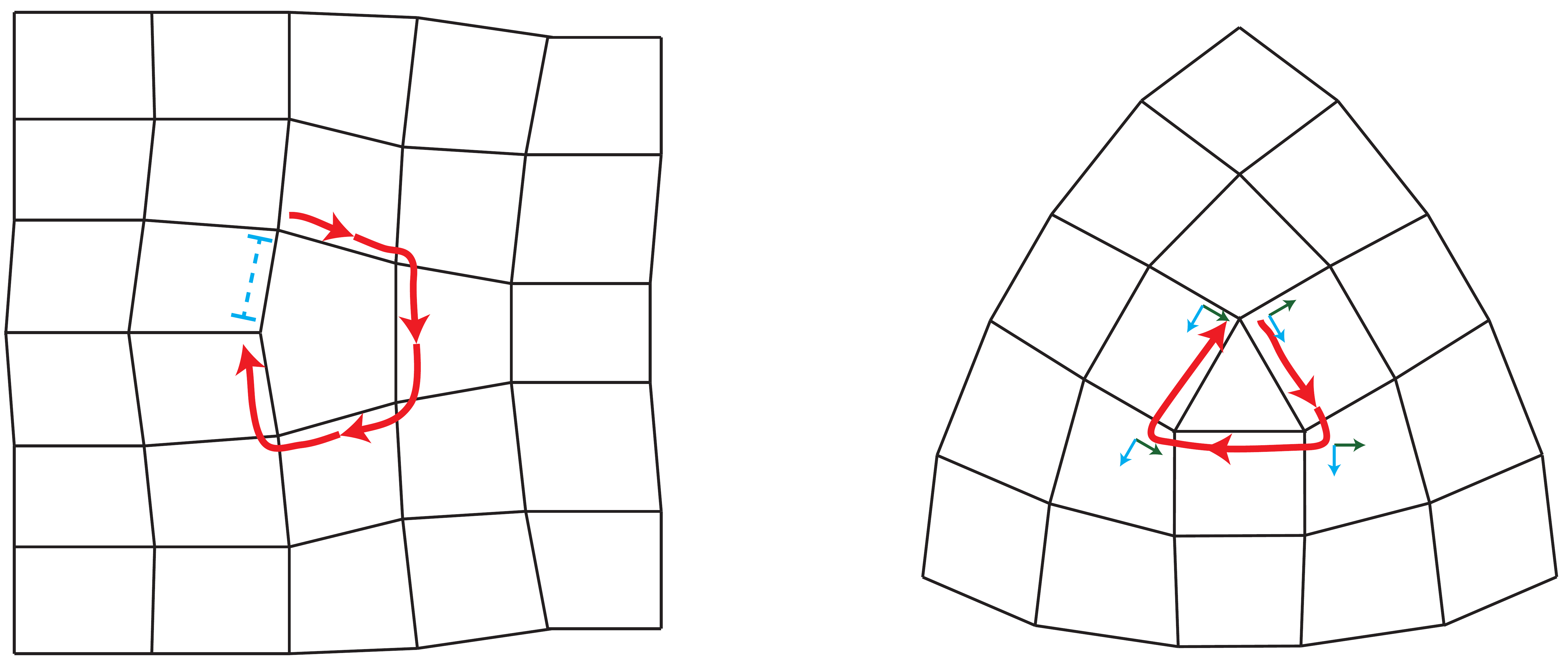}
\caption{Examples of lattice defects in two dimensional lattices, following \cite{Hughes:2012vg}. On the left, we can see a dislocation defect and a non-zero Burgers vector in blue. On the right we have a disclination defect which can be seen from the parallel transport of the frame around the loop. 
\label{fig:lattice_defects}
}
\end{figure}

To motivate a condensed matter interest in geometry, we appeal to the study of lattice defects (see \cite{landau1986theory,lubensky1995principles}). There are two standard types of defects: dislocations and disclinations. One begins with an orthonormal basis of one-forms at every lattice site defining a local   coframe $e^a_\mu = \delta^a_\mu + \partial u^a/\partial x^\mu$. The second term is the distortion tensor, whose symmetric part is the strain. The coframes define a spatial metric and volume form,
\eq{
\delta_{ab} e^a_\mu e^b_\nu = g_{\mu\nu}, \qquad \qquad \det e =\frac{1}{D!}\ce_{ab \ldots c} e^a \wedge e^b \wedge \ldots \wedge e^c.\label{eq:coframe_metric}
}
From the distortion tensor we can measure the Burgers vector around a loop $\cC = \partial \Sigma$,
\eq{
\int_{\partial \Sigma} e^a = - b^a. \label{eq:burgers}
}
The Burgers vector keeps track of dislocations but we also need to keep track of disclinations, which distort the angles in the lattice. To do so, we introduce a connection for rotations, $\omega_\mu{}^a{}_b$, which is  an antisymmetric matrix valued one-form. The Frank angle around a loop is similarly measured,
\eq{
\int_{\partial \Sigma} \omega^a{}_b = - \theta^a{}_b. \label{eq:frank}
}
See figure \ref{fig:lattice_defects} for two-dimensional examples of dislocations and disclinations.

When we make the construction fully rotationally covariant, we require under local frame rotations (which should not be confused with rotations of the coordinates),
\eq{
e^{a'} = \Lambda^{a'}{}_ae^a,\:\, \qquad \qquad \omega^{a'}{}_{b'} = \Lambda^{a'}{}_a \Lambda^{b}{}_{b'} \omega{}^a{}_b - \Lambda^c{}_{b'} d\Lambda^{a'}{}_c.\label{eq:coframe_connection_transform}
}
Any object in a vector or tensor representation of local frame rotations will transform homogeneously as the coframe does. It is a quick exercise to confirm that the curvature $R^a{}_b$ and torsion $T^a$ both transform homogeneously
\eq{
R^a{}_b = d\omega^a{}_b + \omega^a{}_c \wedge \omega^c{}_b,
\qquad \qquad 
T^a = de^a + \omega^a{}_b \wedge e^b.
}
Note that ignoring the quadratic terms we can use Stokes' theorem to relate these two-forms to the  densities that source the Burgers and Frank vectors in (\ref{eq:burgers},~\ref{eq:frank}). We therefore can physically interpret curvature and torsion as covariantized sources for disclinations and dislocations respectively. We note that throughout this paper we define \emph{flat} space to have zero curvature as well as zero torsion.

Now, if we have a matter field in a vector representation, we can use the connection to define a rotationally covariant derivative,
\eq{
v^{a'} = \Lambda^{a'}{}_av^a,
\qquad \qquad 
D_\mu v^a = \partial_\mu v^a + \omega_\mu{}^a{}_b v^b ,
\qquad \qquad 
 D_\mu v^{a'} = \Lambda^{a'}{}_a D_\mu v^{a}.
} 
When we have spinor fields, the spin covariant derivative involves the spinor representation of rotations $(S^{ab})^\a{}_\b = \frac i4 [\g^a, \g^b]^\a{}_\b$, where $\a$ and $\b$ are spinor indices about which we will be rather blas\'e
\eq{
D\psi^\a = d\psi^\a + \frac i2 \omega_{ab} (S^{ab})^\a{}_\b \psi^\b.
}

All of this story can be lifted from a Euclidean rotation group to the Lorentz group if we wish to consider Lorentz invariant theories by allowing both the group index $a$ and the spacetime index $\mu$ to have a zero (time) component. The only subtlety is that we have to keep track of signs as we raise and lower group indices with $\eta^{ab} = \diag{-1,+1,\ldots}$. Such a construction will provide us with theories that are invariant under both boosts and rotations.

An important fact that we wish to emphasize is that for a choice of coframe, there exists a unique spin connection, the Levi-Civita connection $\tilde \omega$ for which the torsion vanishes,
\eq{de^a + \tilde \omega^a{}_b e^b = 0.
}
Further, it immediately follows from \eqref{eq:coframe_connection_transform} that the difference of two spin connections transforms homogeneously. Given a coframe and connection one can define the cotorsion, the difference between the connection supplied and the Levi-Civita connection derived from the coframes,
\eq{
C^a{}_b = \omega^a{}_b - \tilde \omega^a{}_b.}
Note that cotorsion is algebraically related to the torsion,
\eq{
T^a = C^a{}_b\wedge e^b.~
}

\subsection*{Torsion and response}

 When defining continuum theories coupled to torsion, we are left with an arbitrary choice  - we may write kinetic terms in terms of the torsionful or torsionless connections, and the difference will simply be a direct coupling to torsion. If we consider a generic renormalization group flow from a lattice model to a continuum theory, there is no symmetry to prevent either term from being generated (unless we can impose a symmetry that the theory explicitly does not feel torsion whatsoever), and so we expect generic coupling to torsion.

Now that we can couple matter fields to both the coframe and connection, we need to define operators corresponding to response to curvature and torsion. In metric theories with no torsion, the stress tensor is defined via
\begin{equation}
\label{eq:T_flat}
T^{\mu\nu}=\frac{2}{\sqrt{-g}}\frac{\delta S}{\delta g_{\mu\nu}} .
\end{equation}
Since the metric is defined uniquely from the coframe, it is reasonable to define the stress tensor as response to changing the coframe $\delta S/\delta e^a_\mu,$ but we have to choose what we hold fixed as an independent variable: the full connection $\w$  or the cotorsion $C$. If we insist that our theory is a \emph{generalization} of a torsionless metric theory, its responses must reduce to those of said theory when restricted to torsionless backgrounds. This is \emph{not} the case if we take the coframe and connection to be independent since a coframe variation at fixed connection introduces torsion. For example, the stress tensor defined as $\delta S/\delta e^a_\mu$ while keeping the connection $\omega$ fixed does not reduce to \eqref{eq:T_flat} even when calculated on a background without torsion. Therefore throughout this paper we will define the stress tensor $t_a{}^\mu$ and torsion current $s^{\mu ab}$ via
\eq{
\delta_{e,C}S = \int \det e \left[t_a{}^\mu \delta e^a_\mu + \delta C_{\mu ab} s^{\mu ab} \right]. \label{eq:stress_tors_defn}
}
We can now test whether \eqref{eq:hall_spin} holds with orbital spin defined as torsion response,
\eq{
\vev{\ell} = \vev{s^t{}_{12}} = 2\eta_H,\label{eq:hall_spin_tors}
}
with $s^t{}_{12}$ defined as in \eqref{eq:stress_tors_defn}.

\section{Free Dirac fields}
\label{sec:dirac_R}

In this section we calculate the torsion current of a free Dirac field in curved space. There is an ambiguity in the coupling of fermions to torsionful geometries \cite{Nieh:1981xk} and so the usual minimal coupling procedure is fine-tuned and insufficient to capture general dynamics. Beginning with the most general free action we find that the relationship (\ref{eq:hall_spin_tors}) is violated precisely by the new torsional couplings that are allowed. The torsion current  does not correspond to the spin angular momentum of a theory and picks up additional contributions from the theory's response to torsional perturbations. This is to be expected from its definition $\frac{\delta S}{\delta {C_\mu}^{ab}}$ and the disagreement is present even in flat backgrounds. 

The most general action for free Dirac spinors is given in \cite{Hughes:2012vg}
\begin{align}\label{dirac action}
	S = \int \det e\left( \frac{i}{2} \bar \psi \gamma^a \overset{\leftrightarrow} D_a \psi  - m \bar \psi \psi - \frac{i}{8} \alpha C_{abc} \bar \psi \{ \gamma^a , [ \gamma^b , \gamma^c ] \} \psi - \frac{1}{8} \beta C_{abc} \bar \psi [ \gamma^a , [ \gamma^b , \gamma^c ] ] \psi \right) 
\end{align}
where $\alpha$ and $\beta$ are free parameters.
Here the $\g^a$ are $d+1$ matrices satisfying the Clifford algebra
\begin{align}
	\{ \gamma^a , \gamma^b \} = - 2 \eta^{ab}
\end{align}
and $D_\mu = \partial_\mu - i A_\mu + \frac{i}{2} \omega_{\mu ab} S^{ab}$ is the Lorentz and gauge covariant derivative. In \cite{Hughes:2012vg}, the term proportional to $\beta$ is absorbed into the gauge field $A_\mu$ and thereafter ignored. This however alters the response and hence we will not do so here.

Now, since we are treating the coframe and cotorsion as independent, it's convenient to isolate the torsional dependence by breaking up the kinetic term into a part that depends only on the Levi-Civita connection and a part that depends only on the torsion.
Specializing to $2+1$ dimensions, we can also use the gamma matrix identities
\begin{align}
	&\{ \gamma^a , [ \gamma^b , \gamma^c ] \} = 4 i \varepsilon^{abc} ,
	&&[ \gamma^a , [ \gamma^b , \gamma^c ] ] = 4 (\eta^{ac} \gamma^b - \eta^{ab} \gamma^c) 
\end{align}
 to simplify the action (\ref{dirac action}).  We then have
\begin{align}\label{dirac 2+1}
	S = \int \det e \left( \frac{i}{2} \bar \psi \gamma^a \overset{\leftrightarrow} {\tilde D_a} \psi  - m \bar \psi \psi + \frac{1}{2} \left( \frac{1}{2} + \alpha \right)  C_{abc} \varepsilon^{abc}\bar \psi \psi + \beta {C^b}_{ba} \bar \psi \gamma^a \psi \right) 
\end{align}
where $\tilde D_\mu$ is the Levi-Civita Lorentz covariant derivative. The torsion current is now trivial to calculate
\begin{align}\label{spin current}
		{s^\mu}_{ab} = \frac{1}{2} \left( \frac{1}{2} + \alpha \right) {\varepsilon^\mu}_{ab}\bar \psi \psi +  \beta e^\mu_{[a} \bar \psi \gamma_{b]}\psi 
\end{align}
giving a density that is shifted from the actual spin density $\frac{1}{2} \bar \psi \psi$
\begin{align}
	s= e^0_\mu {s^\mu}_{ab} \varepsilon^{ab} = \left( \frac{1}{2} + \alpha \right) \bar \psi \psi \label{eq:dirac_tors_density}
\end{align}
(here $\varepsilon_{ab}$ is the volume element of the 1-2 plane).
The additional contributions to the torsion current spoil our interpretation of it: it is no longer tied to spin which can be defined purely in flat space without the need to consider generalization to curvature or torsion. 

It is also clear that if we define the stress tensor of the theory by a variation with respect to the coframe keeping cotorsion fixed, the stress would be independent of both the $\alpha$ and $\beta$ terms in the action when set on a torsionless background:
\begin{equation}
	T_{\mu\nu}\Big|_{C=0}= \frac i 2 \bar\psi \, \gamma_{(\mu} \overset{\leftrightarrow} {\tilde D}{}_{\nu)}\psi+ \left( \frac{i}{2} \bar \psi \gamma^a \overset{\leftrightarrow} {\tilde D_a} \psi  - m \bar \psi \psi  \right) g_{\mu\nu} . \footnote{The direct variation of the Levi-Civita connection with respect to the metric does not contribute to the symmetric stress.}
\end{equation}
As such, the Hall viscosity cannot depend on these coefficients.

In this theory the effective action can be calculated by integrating out the fermions. Among the terms generated in this way is the gravitational Chern-Simons term \cite{Vuorio:1986ju},  which does not contribute to the Hall viscosity  as it is third order in derivatives of the metric.  In fact, in the absence of a magnetic field no term is generated at the order which can contribute to Hall viscosity.

We note that the detailed Hall viscosity calculation given in \cite{Hughes:2012vg} is consistent with our response functions in that it uses the definition of the stress tensor as the variation of the torsion-free action with respect to the frame, which as argued in section \ref{sec:tors_bg} is equivalent to the variation of the full action while keeping the cotorsion fixed. It was also shown in \cite{Hughes:2012vg} that this is equivalent to the calculation of Hall viscosity from an adiabatic approach, the result being that there is a contribution of the form $e^a\wedge de_a$ to Hall viscosity which also needs to be regulated.

This implies that: 1. Since the microscopic derivation was carried out without any reference to torsion, if there is a non-zero Hall viscosity the EFT should also be able to describe  it consistently without torsion. This means that  there should be no need to reference torsion to make the effective action covariant. However there is no  covariantization of $e^a \wedge de_a$ other than $e^a\wedge T_a$. 2. The calculation of the Hall viscosity from the EFT must be consistent with its calculation from the microscopic theory. That is, it must come from a stress-stress correlator defined by varying the frame while keeping the torsion fixed. As such, the claim in \cite{Hughes:2012vg} that the $e^a \wedge T_a$ term contributes to Hall viscosity is inconsistent\footnote{We also note that in \cite{Hughes:2012vg} the contact term $\vev{\frac{\delta T^{ij}}{\delta g_{j\ell}}}$ was ignored, which \cite{Bradlyn:2012ea} argues contributes to the  viscosity tensor. From  \eqref{dirac 2+1} we find the contact term relevant for Hall viscosity is $
T^{xx} = \frac{1}{4}\psib \psi \dot{h}_{xy}+\cdots $,
where we have dropped terms which do not contribute to Hall viscosity. }. We conclude that the only sensible regulation of the Hall viscosity calculation is to set it to zero. As we will show in the next section, a relativistic gapped system cannot have a non-zero Hall viscosity in a Lorentz invariant background.

\section{Relativistic Quantum Hall}
\label{sec:QH_R}

In this section we analyze the most general effective theory of a relativistic quantum Hall system in the first order formalism. We follow the same line of reasoning as \cite{Golkar:2014,Golkar:2014paa} and generalize to the case of non-zero torsion.

\subsection*{The Euler current with torsion}
As in \cite{Golkar:2014,Golkar:2014paa}, the discussion starts with the definition of the Euler current,  the relativistic analog of the spatial curvature in the Wen-Zee term $A\wedge R$, which now needs to be generalized to the first order formalism. Along with a coframe and a connection, if we are given a globally well defined time-like vector of unit norm $u^\mu$, we can  construct a two form $\Theta$:
\eq{
\Theta[u,e,\omega]=\frac{1}{8 \pi}\epsilon_{abc} u^a \left( Du^b \wedge Du^c -  R^{bc}\right),\label{eq:theta_defn}
}
where $u^a = e^a_\mu u^\mu$ and the covariant exterior derivative is defined with the connection $\omega$, i.e. $Du^a=du^a+{\omega^a}_b u^b$. In this language, the statement that the Euler current is conserved becomes $d\,\Theta = 0.$\footnote{In terms of the notation of \cite{Golkar:2014,Golkar:2014paa} we have $\Theta_{\mu \nu} = (*J)_{\mu\nu}$.} The crucial property of the current is that given a two dimensional spacial surface $\Sigma$, the total charge associated with the Euler current is the Euler character of $\Sigma$ (see \cite{Golkar:2014paa} for discussions and proof):
\eq{
\int_\Sigma \Theta = \chi(\Sigma)/2.\label{eq:euler_charge}
}

Note that the definition of the Euler current and its properties remain valid for any connection. In fact, with a little bit of algebra one can show that given an $\omega$ and an $\omega'$, the difference between the Euler currents defined with these two connections is a total derivative. Taking $Y=\omega'-\omega$, we know that $Y$, being the difference of two connections, is a one form that transforms as a tensor under Lorentz transformations. We have then have
\eq{
\Theta[u,e,\omega'] =\Theta[u,e,\omega+Y] = \Theta[u,e,\omega]-d\left(\frac{1}{8\pi} \ce_{abc}u^aY^{bc} \right).\label{eq:euler_shift}
} 

This identity is particularly useful here, since it can be used to extract the torsion dependent part of this current. That is, we can use the relation between the connection and the Levi-Civita connection $\omega =\tilde \omega + C$ to write:
\eq{
\Theta[u,e,\omega]= \Theta[u,e,\tilde \omega]-d\left(\frac{1}{8\pi} \ce_{abc}u^a C^{bc} \right).\label{eq:euler_shift_torsion}
}  

\subsection*{The effective action}
We will be expanding about a constant magnetic field with a flat vielbein and no torsion or curvature. In terms of our power counting scheme we have
\begin{align}
	&A = \cO(p^{-1}),&&e^a = \cO(1), &&\w^a{}_b = \cO(p), &&T^a = \cO(p), &&R^a{}_b =\cO(p^2).
\end{align}
This allows us to construct a unit norm timelike vector field $u_\mu$
\eq{
b = \sqrt{F_{\mu\nu}F^{\mu\nu}/2},
\qquad \qquad
u_\mu =\frac{1}{2b} \epsilon_{\mu\nu\rho} F^{\nu\rho}.
}
Note that $d(*b u)=dF=0$ and both $b$ and $u_\mu$ are $\cO(1)$ by the power counting above. 

We now look at all possible gauge invariant terms up to first order in a derivative expansion. At leading order $\cO(p^{-1})$, the only gauge invariant term we can write down is the standard Chern-Simons term: $\frac{\nu}{4\pi}A \wedge F$.
At zeroth order in derivative expansion we can write an arbitrary function $\epsilon(b)$. The local gauge invariant one derivative terms we can construct (with the constraint $u_\mu u^\mu=-1$ and ignoring total derivatives) are:
\begin{align}
&f(b) \epsilon^{\mu\nu\rho} u_\mu \partial_\nu u_\rho,&&g_1(b) \epsilon^{\mu\nu\rho} u_\mu T^\lambda_{\nu\rho} u_\lambda,\nonumber \\&g_2(b)\eta_{ab}T^a\wedge e^b,&&g_3(b)\ce_{abc}u^a T^b\wedge e^c.
\end{align}
 We can of course also consider a coupling to the Euler current, $\kappa~ A \wedge \Theta.$ The Lagrangian density is therefore
\begin{multline}
\cL =
\frac{\nu}{4\pi}A \wedge F+ (\det e) \Big[- \epsilon(b) +f(b) \epsilon^{\mu\nu\rho} u_\mu \partial_\nu u_\rho + g_1(b)\, \epsilon^{\mu\nu\rho} u_\mu T^\lambda_{\nu\rho} u_\lambda\Big] \\ 
+g_2(b)\,\eta_{ab}T^a\wedge e^b+g_3(b)\,\ce_{abc}u^a T^b\wedge e^c+\frac{\kappa}{8 \pi}\,A \wedge\ce_{abc} u^a \left( Du^b \wedge Du^c -  R^{bc}\right).\label{eq:tors_action}
\end{multline}
Here, we have chosen to write the Euler current as a function of the connection $\omega$. If we had written the coupling to the Euler current with the Levi-Civita connection $\tilde \omega$, using \eqref{eq:euler_shift} and ignoring boundary terms we would have:
\begin{align}
\frac\kappa{8\pi} \int A \wedge (\Theta[\tilde \omega +C]-\Theta[\tilde \omega])=&-\frac{\kappa }{8\pi}\int A \wedge d(\ce_{abc}u^a C^{bc}) =-\int \frac{\kappa}{8\pi} F \wedge \ce_{abc}u^a C^{bc}\notag\\
=&\int \det e\left(-\frac{\kappa b}{8 \pi} \right) \epsilon^{\mu\nu\rho} u_\mu T^\lambda_{\nu\rho} u_\lambda -\int\frac{\kappa b}{8\pi} \eta_{ab}T^a\wedge e^b,
\label{eq:theta_shift}
\end{align}
where we have used $\e^{\mu\nu\rho}=-3 u^\alpha  u^{[\mu}{\e_\alpha}^{\nu\rho]}$ and $u^2=-1$. This means that if we instead consider  in our effective action the coupling to the Levi-Civita Euler current $\frac{\kappa}{8\pi} \int A \wedge \Theta[\tilde \omega]$ it would be equivalent to \eqref{eq:tors_action} under $g_{1,2}(b) \rightarrow g_{1,2}(b)-\frac{\kappa b}{8\pi}.$ 

Since the only terms that depend on $A$ explicitly and not through $F$ are the Chern-Simons and Euler current coupling terms, we know the total charge depends only on those terms, and all other contributions to the charge density are total derivatives. Noting that the Euler character is well-defined even with torsion, we again find:
\eq{
Q = \nu N_\phi+\kappa \frac{\chi}{2}.
}
This can also be seen from equation \eqref{eq:theta_shift}, as $A_\mu$ dependence of the Euler current is precisely the same as that of the Euler current built from the Levi-Civita connection. 
\subsection*{Hall viscosity and response}
When we consider linear response about flat space the Hall viscosity comes only from the Levi-Civita part of the Euler current:
\eq{
\delta^2_{\eta_H} S = -\int \frac{\kappa B}{32\pi} \epsilon^{ij} h_{ik}\dot h_{jk},
\qquad \qquad
\eta_H = \frac{\kappa B}{8 \pi}.
}
There is no Lorentz invariant term which gives Hall viscosity built only out of geometric data. This implies that any relativistic gapped phase  must have vanishing Hall viscosity in the Lorentz invariant vacuum (e.g. vanishing background electromagnetic fields).

We now calculate various one point correlation functions. First consider response to a static magnetic field. The standard current and stress response is \cite{Golkar:2014,Golkar:2014paa}
\eqn{
j^t&=&\frac{\nu}{2\pi}B+\frac{\partial_i^2f(B)}{B}-\frac{f'(B)}{B^2} (\partial_i B)^2,
\qquad
\qquad
j^i =-\epsilon^{ij} \partial_j \ce'(B),\nonumber \\
T^{tt}&=&\ce(B),
\qquad
\qquad
T^{ii}=B \ce'(B)-\ce(B),
\qquad
\qquad
T^{ti}=-\epsilon^{ij} \partial_j \left(\frac{\kappa B}{8\pi}+f(B) \right).
}
We can  also calculate the torsion current,
\begin{align}
	&s^t{}_{12}=\frac{\kappa B}{8\pi}-g_2(B),
	&&s^x{}_{01}= s^y{}_{02}=\frac{g_3(B)}{2}, \nonumber \\
	&s^y{}_{01}=-s^x{}_{02}=g_2(B)-g_1(B).
\label{eq:mag_resp}
\end{align}
Note that the term $g_2(B) \eta_{ab} T^a \wedge e^b$ shifts the torsion current density, which means we can no longer interpret it as  the orbital spin density related to Hall viscosity, similar to what was noted in the non-relativistic case \cite{Bradlyn:2014wla}. If we consider response to $T^a$ instead of the cotorsion, we find 
\eq{
\frac{\delta S}{\delta T^0_{xy}} = - \frac{\kappa B}{8\pi} -g_2(B)+2 g_1(B).
}
Since our background is flat space, the stress and current results are the same as in \cite{Golkar:2014paa}. If we turn on a (perturbative compared to $B$) inhomogeneous but static electric field, we get in addition to \eqref{eq:mag_resp} (which does not change at this order),
\begin{align}
	s^t{}_{0i} &=-\frac{\kappa E^i}{8\pi}+ \frac{E^i g_1(B)+\epsilon^{ij} E^j g_3(B)/2}{B}, \nonumber \\
	s^i{}_{12}&=\frac{\epsilon^{ij} E^j}{8\pi}-\frac{E^i g_3(B)/2+\epsilon^{ij} E^j g_1(B)}{B}.
\end{align}

To the order we work our action is linear in cotorsion and therefore the spin-spin correlation vanishes.

\section{Free Non-relativistic fields}
\label{sec:NR_free}

Finally, we point out that the above considerations are also relevant for the non-relativistic case in an example along the lines of section \ref{sec:dirac_R}. Namely, the most general non-relativistic microscopic theory will couple to torsion in a manner that shifts the torsion current so that it is not directly tied to spin. We need to be sure that none of the terms we introduce break the symmetries of non-relativistic theories and so we employ the formalism of Newton-Cartan geometry. This formalism was first introduced by Cartan in 1923 to present non-relativistic physics in a manifestly coordinate independent fashion \cite{Cartan:1923zea,Cartan:1924yea} and was shown by K{\"u}nzle in 1972 to be the natural structure preserved by Galilean group \cite{Kuenzle:1972zw} (recently enlarged to the Bargmann group in \cite{Brauner:2014jaa}). The subject has undergone  a revival in recent years but for our purposes we particularly note the importance of Milne invariance \cite{Jensen:2014aia} \cite{Hartnog:2014rf}. The interested reader may also refer to \cite{Jensen:2014aia} for a more complete list of references regarding the history and application of Newton-Cartan geometry.

The basic elements of Newton-Cartan geometry are a ``clock form'' $n_\mu$, a degenerate but positive semi-definite ``metric'' $g^{\mu \nu}$ whose kernel is spanned by $n_\mu$, as well as a spacetime connection $\nabla_\mu$ that annihilates both
\begin{align}
	\nabla_\mu n_\nu = 0
	&&\nabla_\lambda g^{\mu \nu} = 0 .
\end{align}
The clock form defines a notion of elapsed time and a preferred spatial slicing so long as $n \wedge dn \neq 0$, which we shall always require. $g^{\mu \nu}$ then defines a Riemannian metric on these slices. The connection's compatibility with $n_\mu$ implies that the manifold will not in general be torsion free; rather it's temporal torsion must satisfy
\begin{align}\label{timelike torsion}
	n_\lambda {T^\lambda}_{\mu \nu} = (d n )_{\mu \nu} .
\end{align}
Usually one assumes that this is the only torsion present. Of course, for our purposes we will take ${T^\lambda}_{\mu \nu}$ to be arbitrary, subject only to the condition (\ref{timelike torsion}).

Since the metric is degenerate, it has no inverse, but it is often convenient to introduce a time-like vector field $v^\mu$ (time-like in the sense that $n_\mu v^\mu = 1$) that allows us to form a partial inverse $g_{\mu \nu}$. This is uniquely fixed by the conditions
\begin{align}
	g_{\mu \nu} v^\nu = 0,
	&& g_{\mu \lambda} g^{\lambda \nu} = {P_\mu}^\nu 
\end{align}
where ${P_\mu}^\nu = {\delta_\mu}^\nu - n_\mu v^\nu$ is the transverse projector orthogonal to $v_\mu$ and $n^\mu$. Of course there are many possible selections of $v^\mu$ that satisfy the defining relation $n_\mu v^\mu = 1$ and we may redefine it at will subject to this constraint. The authors of \cite{Duval:2008jg,Duval:2009vt} demonstrated that under this shift the vector potential should also transform
\begin{align}
	v^\mu \rightarrow v^\mu + \varphi^\mu ,
	&&A_\mu \rightarrow A_\mu + m \left( \varphi_\mu - \frac{1}{2} n_\mu \varphi^2 \right) .
\end{align}
Here $\varphi^\mu$ is an arbitrary transverse vector $n_\mu \varphi^\mu = 0$ and $m$ the mass of the field that $A_\mu$ couples to. 
This transformation is called a Milne boost and all non-relativistic actions must be invariant under it \cite{Jensen:2014aia}. 
The reader may verify for instance that the free Schr{\"o}dinger action, considered in Eqn. (\ref{torsionless micro}), satisfies this requirement.

Finally, we are concerned with spinful particles, so we also need a non-relativistic vielbein and spin connection. This proceeds mostly as in the pseudo-Riemannian case. Introduce a vielbein $e_a^\mu$, $a=1, \cdots , d$ such that
\begin{align}
	g^{\mu \nu} = e^{a \mu} e_a^\nu .
\end{align}
The vielbein is in the $SO(d)$ fundamental and we will raise and lower $SO(d)$ indices with the metric $\delta^{ab}$ throughout. The spin connection is then defined by
\begin{align}
	\nabla_\mu e^{a\nu} = -{\omega_\mu}^a{}_b e^{b \nu} 
\end{align}
and is valued in $\mathfrak{so}(d)$. In the $2+1$ dimensional case that we are concerned with the spin connection is abelian and it is convenient to work with
\begin{align}
	\omega_\mu = \frac{1}{2} \varepsilon_{ab} \omega_\mu{}^{ab}
\end{align}
instead, which transforms like a $U(1)$ gauge field.

Non-relativistic gapped theories were considered in \cite{Bradlyn:2014wla} where the authors demonstrate the relationship between the Hall viscosity and torsion response by writing the most general local effective action $W[A , e^a , C^{ab}]$ and calculating $T^{\mu \nu}$ and $s^\mu{}_{ab}$. In a crucial step they set explicit torsionful terms 
\begin{align}
	W = \int \det e\left( \gamma (b) \varepsilon^{\mu \nu \lambda} T_{\mu \nu \lambda} + \cdots \right)
\end{align}
to zero, arguing that torsion does not appear in the usual UV theory describing spinless particles coupled to geometry
\begin{align}\label{torsionless micro}
	S = \int \det e \left( \frac{i}{2} v^\mu \psi^\dagger \overset{\leftrightarrow} D_\mu \psi - \frac{1}{2m} g^{\mu \nu} ( D_\mu \psi )^\dagger ( D_\nu \psi )\right) 
	&&\text{with}
	&&D_\mu = \partial_\mu - i A_\mu
\end{align}
and so cannot enter the effective theory. 

However, as in the relativistic case, the most general microscopic theory will couple to torsion and our ignorance of these terms from the flat space point of view should not preclude their appearance in the general theory. Indeed for electrons moving in a background lattice one would expect interference with crystalline defects. Describing such a process in the continuum theory necessitates nontrivial torsional couplings. In this section we find the most general such couplings to a chosen dimension involve coupling the cotorsion to number current and number density. 

We seek all Milne invariant operators of dimension 3 or 4 (the same dimension as what appears in (\ref{torsionless micro})) and involve the cotorsion $C_\mu \equiv \frac{1}{2} \varepsilon_{ab} {C_\mu}^{ab}$ which is itself Milne invariant and of dimension 1.\footnote{ This is in conflict with the work of \cite{Jensen:2014aia} whose connection varies under Milne transformations in a torsionful background. We take the point of view that the connection must be Milne invariant as a notion of parallel transport is physically meaningful and will show in forthcoming work that one can obtain a Milne invariant connection in a torsionful Newton-Cartan setting \cite{TBA_GPR}.}
There is one Milne invariant vector and one Milne invariant scalar of dimension 2
\begin{align}
	\mathcal J^\mu \equiv v^\mu \psi^\dagger \psi - \frac{i}{2m} \psi^\dagger {\overset{\leftrightarrow} D}{}^\mu \psi ,
	&& n_\mu \mathcal J^\mu = | \psi |^2,
\end{align}
which are simply the number current and density. Note that even if the fields are spinless our coupling of torsion to number current is allowed. The most general possible action for spin $s$ fields to our order is then
\begin{align}
	S = \int \det e\left( \frac{i}{2} v^\mu \psi^\dagger \overset{\leftrightarrow} D_\mu\psi - \frac{1}{2m} g^{\mu \nu} (D_\mu \psi )^\dagger (D_\nu \psi ) + \alpha C_\mu \mathcal J^\mu + \frac{1}{2} \beta C_\mu C^\mu | \psi |^2\right) ,
\end{align}
where now $D_\mu = \partial_\mu - i A_\mu- is \omega_\mu$. We calculate the torsion current
\begin{align}
	s^\mu =  ( s + \alpha ) \mathcal J^\mu + \beta C^\mu | \psi |^2
	\qquad
	\implies \qquad
	n_\mu s^\mu = ( s+ \alpha ) | \psi |^2 .
\end{align}
Again, the spin current is shifted by the additional torsion terms while the the stress tensor is independent of the new terms when restricted to a torsionless background. Let's calculate the stress to confirm this. Introduce a variation
\begin{align}
	&\delta e^a_\mu = \frac{1}{2} \delta h_{\mu \nu} e^{a \nu} 
\end{align}
to the vielbein
where $\delta h_{\mu \nu} v^\nu = 0$. This induces a change to the Levi-Civita spin connection
\begin{align}
		\delta \tilde \omega_\mu = - \frac{1}{2} \varepsilon^{\nu \lambda} \tilde \nabla_\nu \delta h_{\mu \lambda} 
\end{align}
where, as before, we denote Levi-Civita objects with a tilde.
Using this to directly vary the action we find on torsionless backgrounds
\begin{align}\label{schrod stress}
	T^{\mu \nu}|_{C=0} 
		= &\left( \frac{i}{2} v^\lambda \psi^\dagger \overset{\leftrightarrow} D_\lambda \psi - \frac{1}{2m} ( D_\lambda \psi )^\dagger ( D^\lambda \psi ) \right) g^{\mu \nu} + \frac{1}{m} ( D^{ (\mu } \psi)^\dagger (D^{\nu)} \psi) \nonumber \\
		&- \frac{1}{2} s \tilde \sigma^{\mu \nu} | \psi |^2 - \frac{is}{2m} \varepsilon^{\lambda ( \mu} \nabla_\lambda \left( \psi^\dagger \overset{\leftrightarrow} D{}^{\nu )}\psi \right).
\end{align}
Here $\varepsilon_{\mu \nu} = {\varepsilon_{\mu \nu \lambda}} v^\lambda$ defines a spatial area form where $\varepsilon_{\mu \nu \lambda}$ is the natural spacetime volume element determined by the connection. (\ref{schrod stress}) also involves the shear tensor
\begin{align}
	\sigma^{\mu \nu} = - \pounds_v g^{\mu \nu} = \nabla^\mu v^\nu + \nabla^\nu v^\mu
\end{align}
and its dual
\begin{align}
	\tilde \sigma^{\mu \nu} = \frac{1}{2} \left( \sigma^{\mu \lambda} {\varepsilon_\lambda}^\nu + \sigma^{\nu \lambda} {\varepsilon_\lambda}^\mu \right) .
\end{align}
There is no dependence on $\alpha$ or $\beta$ and similarly there can be no dependence on these variables in the Hall viscosity.

\section{Conclusion and outlook}
In this paper, we have shown through a number of examples that the torsion response cannot be identified with either the spin density or Hall viscosity. The relationship is generically broken due to the presence of non-universal torsion couplings in the action which are not forbidden by any symmetries. One would expect these terms to be generated from a model on a lattice with defects. It would be instructive to see how this arises in coarse-graining to the continuum limit.

There are several questions immediately motivated by our analysis. First of all, the correct orbital spin density is intuitively clear when we have access to a microscopic theory. Can one define a general spin density that may be applied to EFTs and that reduces to this concept when it is available? Secondly, what is the proper physical interpretation of the torsion current? Though the torsion current and spin density are not the same, is there any relationship between the two? Furthermore, we have argued that a relativistic Hall viscosity may only be obtained from the Euler current. Can one derive the Euler current term by integrating out massive states? We defer the analysis of these questions to future work.

\section*{Acknowledgments}
It is a pleasure to thank  Sungjay Lee, Eun-Gook Moon and Nicholas Read for discussions and  Dam T. Son for insightful discussions and encouragement.  This work is supported in part by DOE grant DE-FG02-13ER41958.
S.G. is supported in part by NSF MRSEC grant DMR-0820054.

\addcontentsline{toc}{section}{Bibliography}
\bibliographystyle{JHEP}
\bibliography{WenZeerefs}

\end{document}